\documentclass[preprints,article,accept,moreauthors,pdftex]{Definitions/mdpi}

\firstpage{1} 
\pubvolume{1}
\issuenum{1}
\articlenumber{0}
\pubyear{2022}
\copyrightyear{2022}
\datereceived{} 
\dateaccepted{} 
\datepublished{} 
\hreflink{https://doi.org/} 
\pdfoutput=1

\Title{TYPE I SHAPOVALOV WAVE SPACETIMES IN THE BRANS-DICKE SCALAR-TENSOR THEORY OF GRAVITY}

\TitleCitation{Type I Shapovalov wave spacetimes in the Brans-Dicke scalar-tensor theory of gravity}

\Author{
Konstantin Osetrin $^{1, 2,
\ddagger}$\orcidA{0000-0003-4739-2788}, 
Altair Filippov$^{1,\ddagger}$ \orcidB{0000-0003-0521-0958},
Ilya Kirnos $^{1,\ddagger}$ \orcidC{0000-0003-3737-3583}, 
and 
Evgeny Osetrin $^{1,\ddagger}$ \orcidD{0000-0001-5998-2238}
}

\AuthorNames{Konstantin Osetrin, Altair Filippov, Ilya Kirnos and Evgeny Osetrin}

\AuthorCitation{Osetrin, K.E.; Filippov A.E.; Kirnos, I.V.; Osetrin, E.K.}

\address{%
$^{1}$ \quad Tomsk State Pedagogical University; osetrin@tspu.edu.ru\\
$^{2}$ \quad Tomsk State University; osetrin@gmail.com}

\secondnote{These authors contributed equally to this work.}

\abstract{
Exact solutions for Shapovalov wave spacetimes of type I in the scalar-tensor theory of gravity of Brans-Dicke are constructed. Shapovalov's wave spacetimes describe gravitational-wave models that allow the separation of wave variables in privileged coordinate systems. In contrast to the general theory of relativity, the vacuum field equations of the Brans-Dicke scalar-tensor theory of gravity lead to exact solutions for type I Shapovalov spaces, which makes it possible to construct observational checks for detecting such wave disturbances. For the models under consideration, equations for the trajectories of test particles are obtained.
}


\begin{document}
%

\section{Introduction}


The recently proposed classification of Shapovalov wave spacetimes \cite{Osetrin2020Symmetry} provides additional mathematical tools for constructing exact integrable models of gravitational waves, including primordial gravitational waves in Bianchi universes \cite{Lukash1976268,Osetrin2022894
}
and other exact models of plane gravitational waves
\cite{
Osetrin2020410,Osetrin2020403}.
Current advances in gravitational wave astronomy in detecting gravitational waves and the resulting astrophysical information
\cite{PhysRevLett.116.061102,PhysRevX.9.031040,PhysRevX.11.021053} increased interest in the mathematical aspects of gravitational wave research as well. 

The relevance of these lines of research is also related to the importance of the discovery of primordial gravitational waves for constructing a theory of the early universe and possible confirmation of the stage of inflation predicted by some theories.
Interest in the study of gravitational waves is also due to possible secondary physical effects in a gravitational wave, such as the formation of black holes \cite{Saito200916,Saito2010867}, the capture of astrophysical objects by a gravitational wave
\cite{Bialynicki-Birula2018}, gravitational wave lensing and other effects.

The Shapovalov wave spacetimes allow the presence of special symmetries of the Hamilton-Jacobi equation, leading to the existence of privileged coordinate systems, where the equations of motion of test particles and the eikonal equation for radiation in the Hamilton-Jacobi formalism allow exact integration by the method of separation of variables. Moreover, among the separable variables on which the space-time metric depends (non-ignored variables),
there are wave variables along which the space-time interval vanishes. Gravitational wave velocity according to gravitational wave detection data from neutron star mergers \cite{Abbott2017PRL161101}
equal to the velocity of light. Therefore, the existence of the possibility of separating null variables indicates the wave nature of such spacetime models, and the use of wave variables in describing gravitational waves is experimentally justified.

Shapovalov wave spaces in the four-dimensional case include three main types according to the number of commuting Killing vectors they allow and, accordingly, in privileged coordinate systems, the number of Killing vectors determines the number of non-ignorable variables on which the space-time metric depends. For privileged coordinate systems, Vladimir Shapovalov obtained the structure of the space metric \cite{Shapovalov1978I,Shapovalov1978II,Shapovalov1979}, which leads to the separation of variables in the Hamilton-Jacobi equation, which allows one to construct exact integrable models of gravitational waves in various theories of gravity, where test particles move along geodesic curves of spacetime.

There is currently interest in studying the properties of modified theories of gravity, which could give corrections to Einstein's theory of gravity in the early stages of the universe (quantum corrections, quadratic theories and 
$f(R)$-theories of gravity, theories with a scalar field, etc.) and could offer theoretical description of the phenomena of ''inflation'', ''dark matter'' and ''dark energy'' 
\cite{Odintsov2007115,Odintsov201159,Odintsov2017104,Osetrin2021092501}.
Therefore, the use of additional mathematical tools for the comparative analysis of exact models of gravitational waves in various theories of gravity also makes it possible to select the most realistic theories and models.

In this direction, one of the first models of modified theories of this kind was the Brans-Dicke scalar-tensor theory of gravity \cite{Brans1961925}. Scalar-tensor theories have firmly taken their place in theoretical research on gravity and cosmology \cite{Novikov2014352}, which is determined by attempts to use an additional scalar field to describe various possible scenarios for the dynamics of the universe and other theoretical constructions
\cite{Odintsov2022729,Odintsov2022100950,Odintsov2022136817,Capozziello2021100867}.

In this paper, Shapovalov wave spaces of type I are considered in the scalar-tensor theory of gravity of Brans-Dicke. An additional scalar field in theory complicates the field equations, but also provides additional possibilities. The resulting equation for a scalar field of the type of the Klein-Gordon equation also requires additional study. As Shapovalov \cite{Shapovalov1978I,Shapovalov1978II} showed, the scalar equation can be integrated by separation of variables in the same privileged coordinate systems as the Hamilton-Jacobi equation. Therefore, the use of Shapovalov wave spaces also provides additional possibilities for the exact solution of the scalar equation \cite{Osetrin20181383,Osetrin20202050275,Osetrin20202050184, Osetrin20211173}. Recently, in the study of the symmetries of the {Klein}-{Gordon}-{Fock} equations, new results have also been obtained
\cite{Obukhov2021727,Obukhov202113201320,Obukhov2022346,Obukhov2022023505}.

Type I Shapovalov metrics are the most general gravitational-wave models of Shapovalov spacetimes, since they depend on three variables in the privileged coordinate system, including the wave variable. Note that metrics of this type when using Einstein's vacuum equations lead to degeneration, since the number of non-ignorable variables in the metric decreases when solving vacuum field equations and the space becomes either a type II wave Shapovalov space with two non-ignorable variables in the metric, or a type III space with one wave variable in metric. Preservation of type I metrics is possible if there are additional fields and matter in the models under consideration. Therefore, the existence of non-degenerate Shapovalov models of type I in the Brans-Dicke theory of gravity and the possibility of detecting such gravitational waves in observations would be additional evidence in favor of scalar-tensor theories of gravity.

The authors investigate exact wave models of spacetime to find out the differences that may arise in these gravitational wave models for Einstein's theory and for modified theories of gravity. Type I Shapovalov wave spaces have internal symmetries that relate them to mathematical models of gravitational waves, but, nevertheless, these models of gravitational waves do not lead to the appearance of exact solutions for Einstein's vacuum equations, in contrast to type II and III models. And in the modified Brans-Dicke theory of gravity, exact wave solutions for given spaces, as we show below, arise. This makes it possible to obtain observational differences in the detection of gravitational waves by existing and future detectors. This is also significant in the analysis of stochastic gravitational wave noise detected by existing gravitational wave detectors. In addition, primordial gravitational waves of this type could leave an imprint on the cosmic microwave background. Finding "traces" of such gravitational waves would argue for greater realism in scalar-tensor gravity theories than Einstein's theory. The possibility of exact integration of the equations of motion of test particles and the geodesic deviation equations in these models of gravitational waves allows gravitational wave detectors to more efficiently analyze and select such disturbances in signals with high noise.

 \section{
 Gravitational-wave spacetimes of Shapovalov
 }
 
Shapovalov wave spaces allow exact integration of the equations of motion of test particles in the Hamilton-Jacobi formalism
\begin{equation}
g^{\alpha\beta}\frac{\partial S}{\partial x^\alpha}\frac{\partial S}{\partial x^\beta}=m^2c^2
,\qquad
\alpha,\beta,\gamma=0,1,2,3.
\label{HJE}
\end{equation}
with the separation of non-ignorable wave variables (the space metric depends on these variables), along which the space-time interval vanishes \cite{Osetrin2020Symmetry}.
Here $S$ is the action function of the test particle, $m$ is the mass of the particle, $c$ is the speed of light. Let us choose a system of units in the future, where the speed of light is equal to unity.

Type I Shapovalov spacetimes admit one Killing vector, which is included in the so-called ''complete set'' of vectors and Killing tensors of the second rank, which determine integrals of motion linear and quadratic in momenta. In a privileged coordinate system, where the equation (\ref{HJE}) allows complete separation of variables, the type I Shapovalov space-time metric depends on three variables, including the null wave variable, which will be denoted by $x^0$. Another null variable $x^1$ is ignored (cyclic) and is not included in the metric.

Let us consider a model with a type I Shapovalov wave space metric, which in a privileged coordinate system can be written in the following form:
\begin{equation} 
g^{\alpha\beta}=\frac{1}{f_0}
\left(
\begin{array}{cccc}
0&1&0&0\\
1&0&0&0\\
0&0&\frac{1}{W}&0\\
0&0&0&\frac{1}{W}
\end{array}
\right)
\label{metric1}
,\end{equation}
where
\begin{equation}
f_0=f_0(x^0),\ \ \ \  W(x^2,x^3)=t_3(x^3)-t_2(x^2)
.\end{equation}

The scalar curvature ${R}$  takes the following form
\begin{equation}
{R}(x^0,x^2,x^3) = 
\frac{{f_0} 
}{({t_2} - {t_3})^3}
\Bigl(
({t_2} - {t_3}) \left({t_2}'' - {t_3}''\right)-{{t_2}'}^2-{{t_3}'}^2
\Bigr)
,
\end{equation}
where the top prime means the ordinary derivative with respect to the variable on which the function depends.

The nonzero components of the Riemann curvature tensor have the following form:
\begin{equation}
{R}_{0202} = {R}_{0303} =\frac{\left({f_0}'{}^2-2 {f_0}{} {f_0}''\right) ({t_2} - {t_3})}{4 {f_0}{}^3}
,\end{equation}
\begin{equation}
{R}_{2323} = \frac{({t_2} - {t_3}) \left({t_2}'' - {t_3}''\right)-{t_2}'{}^2-{t_3}'{}^2}{2 {f_0}{} ({t_2} - {t_3})}
\label{riman}
.\end{equation}
In addition, all nonzero components of the Weyl conformal curvature tensor $C_{\alpha\beta\gamma\delta}$ are proportional to the scalar curvature $R$
and when the scalar curvature vanishes, they also vanish, which leads to a conformally flat spacetime.

Note that the ability to exactly integrate the Hamilton-Jacobi equations for Shapovalov wave spaces allows one to obtain the trajectories of test particles, as well as to find exact solutions to the geodesic deviation equation and the exact form of tidal accelerations in these spaces. These possibilities make it possible to determine all physical effects in gravitational waves.

\section{
Brans-Dicke scalar-tensor theory of gravity
}

The Lagrange function for the Brans-Dicke theory of gravity (BDT) can be written in the following form \cite{Novikov2014352}:
 \begin{equation}
L=\phi\, (R+2\Lambda)-\frac{\omega}{\phi}\, g^{\alpha\beta}\partial_{\alpha}\phi\partial_{\beta}\phi+16\pi\kappa L_{matter}
,
\end{equation}
where $\phi$ is a scalar field, $\omega$ is a constant parameter of the theory, $\Lambda$ is a cosmological constant.

Field equations of the Brans-Dicke scalar-tensor gravity theory with cosmological constant
$\Lambda$ can be written as follows:
\begin{equation} 
G_{\alpha\beta}=
\frac{8\pi}\Phi T_{\alpha\beta}+\Lambda g_{\alpha\beta}+\frac{\omega}{\Phi^2}(\Phi_{,\alpha}\Phi_{,\beta}-\frac 12 g_{\alpha\beta}g^{\gamma\delta}\Phi_{,\gamma}\Phi_{,\delta})+
	\frac 1\Phi(\Phi_{;{\alpha\beta}}-g_{\alpha\beta} \Phi^{;\gamma}{}_{;\gamma})
\label{EqBD}
,\end{equation}
\begin{equation}
\frac{3+2\omega}\Phi \,\Phi^{;\gamma}{}_{;\gamma}
=
\frac{8\pi}\Phi T^\gamma{}_\gamma+2\Lambda
\label{EqScalar}
.\end{equation}
Here, the comma denotes the ordinary partial derivative, and the semicolon denotes the covariant derivative.

To obtain exact gravitational wave solutions, consider the vacuum model
$T_{\alpha\beta}=0$. It is convenient to pass to the form of the scalar field $\Phi=e^\phi$, then the BDT equations will take the following more compact form:
\begin{equation} 
G_{\alpha\beta}+g_{\alpha\beta}(\phi^{;\gamma}{}_{;\gamma}+(1+\omega/2)\phi^{,\gamma}\phi_{,\gamma}-\Lambda)-\phi_{;{\alpha\beta}}-(\omega+1)\phi_{,\alpha}\phi_{,\beta}=0
,\label{tbd}
\end{equation}
\begin{equation}
(\omega+3/2)(\phi^{;\gamma}{}_{;\gamma}+\phi^{,\gamma}\phi_{,\gamma})-\Lambda=0
.\label{trac}
\end{equation}
Note that the higher derivatives of the scalar field $\phi$ can be found from the equations (\ref{tbd}) and substituted into the scalar equation (\ref{trac}) to reduce the order of this equation.

The equations of the scalar-tensor theory of gravity of Brans-Dicke for the considered metric lead to the following form of field equations:
\begin{equation} \phi_{,00}=\frac{f_0{}'}{f_0}\phi_{,0}+\frac 32\frac{{f_0{}'}^2}{f_0{}^2}-(\omega+1)\phi_{,0}{}^2+\frac{f_0{}''}{f_0},\end{equation}
\begin{equation} \phi_{,01}=(\omega+1)\left(\frac{2\Lambda f_0}{3+2\omega}-\phi_{,0}\phi_{,1}\right),\end{equation}
\begin{equation} \phi_{,11}=-(\omega+1)\phi_{,1}{}^2,\end{equation}
\begin{equation} \phi_{,1\mu}=-(\omega+1)\phi_{,1}\phi_{,\mu},
\qquad \mu,\nu=2,3,\end{equation}
\begin{equation} \phi_{,0\mu}=\phi_{,\mu}\left(\frac 12\frac{f_0{}'}{f_0}-(\omega+1)\phi_{,0}\right),\end{equation}
$$
\phi_{,\mu\mu}=W\left(\frac 13(R+\Lambda)f_0-\frac 12\frac{f_0{}'}{f_0}\phi_{,1}+\frac 13\omega\phi_{,0}\phi_{,1}\right)-\frac{5\omega+6}6{\phi_{,2}}^2
$$
\begin{equation} 
\mbox{}
+\omega
  {\phi_{,3}}^2-(-1)^{\mu}\frac{\phi_{,2}t_2{}'+\phi_{,3}t_3{}'}{2W},\end{equation}
\begin{equation} \phi_{,23}=\frac{\phi_{,2}t_3{}'+\phi_{,3}t_2{}'}{2W}-(\omega+1)\phi_{,2}\phi_{,3},\end{equation}
\begin{equation} 
W((2\omega+3)R+2(4\omega+3)\Lambda)f_0=\omega(2\omega+3)(2W\phi_{,0}\phi_{,1}+{\phi_{,2}}^2+{\phi_{,3}}^2)
.
\label{EqSovm0}
\end{equation}
The equation (\ref{EqSovm0}) arises as a consequence of the scalar equation (\ref{EqScalar}) when the higher derivatives of the scalar field $\phi$ from the other field equations of the theory are substituted into it.

\subsection{Scalar field of the form 
$\phi=\phi(x^0,x^1,x^2,x^3)$ and $\phi_{,1}\neq 0$.}

Assume that the scalar field $\phi$ depends on all variables, including the null variable $x^1$, on which the spacetime metric does not depend.

In this case, the system of BDT equations is quite cumbersome, but it allows one to find the conditions for the compatibility of field equations, which leads to relations of the following form:
\begin{equation} \phi_{,2}(2f_0f_0{}''-3{f_0{}'}^2)=\phi_{,3}(2f_0f_0{}''-3{f_0{}'}^2)=0
\label{sovm2}
.\end{equation}

The requirement $\phi_{,1}\neq 0$ imposes strict conditions on the dependence of the scalar field on the variables $x^2$ and $x^3$, so the case of ${\phi_{,2}}^2+{ \phi_{,3}}^2\neq 0$ in (\ref{sovm2})
leads to the degeneration of the space into a flat one for any $\omega$.

Thus, the scalar field for $\phi_{,1}\ne 0$ can only have the following form:
\[ \phi=\phi(x^0,x^1).\]
In this case, the system of BDT equations takes the following form:
\begin{equation} \phi_{,00}=\frac{f_0{}'}{f_0}\phi_{,0}+\frac 32\frac{{f_0{}'}^2}{f_0{}^2}-(\omega+1)\phi_{,0}{}^2+\frac{f_0{}''}{f_0}\end{equation}
\begin{equation} \phi_{,01}=\frac{\Lambda}2f_0-\frac{\omega+2}2\phi_{,0}\phi_{,1}-\frac 14\frac{f_0{}'}{f_0}\phi_{,1}\end{equation}
\begin{equation} \phi_{,11}=-(\omega+1)\phi_{,1}{}^2\label{eqq11}\end{equation}
\begin{equation} (2\omega f_0\phi_{,0}-f_0{}')(2\omega+3)\phi_{,1}-2(2\omega+1)\Lambda f_0{}^2=0\end{equation}
\begin{equation} (R+\Lambda)f_0+\omega\phi_{,0}\phi_{,1}-\frac 32\frac{f_0{}'}{f_0}\phi_{,0}=0\end{equation}
Consider the equation(\ref{eqq11}). For $\omega\neq -1$ the scalar field is in the form
\begin{equation}
\phi(x^0,x^1)=\frac 1{\omega+1}\ln{\Bigl(a(x^0)x^1+b(x^0)\Bigr)}
,
\end{equation}
however, as a result of substitution into other equations of the system and separation of the coefficients of $x^1$, it turns out that $a(x^0)=0$, and the scalar field does not depend on $x^1$. Consequently,
a solution that satisfies the stated requirements is possible only for $\omega=-1$.

Integration of the equation (\ref{eqq11}) and separation of the coefficients for $x^1$ for $\omega=-1$  leads to the scalar field $\phi=\alpha x^1+b(x^0)$, $ \alpha - const$.
Substitution
into other equations of the system allows to find all unknown functions.

The final solution for the case $\phi=\phi(x^0,x^1)$  is as follows:
\begin{equation}
f_0=e^{\beta x^0}
,\end{equation}
\begin{equation}
\left({t_{2}}' \right)^2=2\alpha\beta\, t_{2}{}^3+\lambda\, t_{2}{}^2
+\gamma\, t_{2}+\delta 
,
\end{equation}
\begin{equation}
\left({t_{3}}' \right)^2=-\Bigl(2\alpha\beta\, t_{3}{}^3+\lambda\, t_{3}{}^2
+\gamma\, t_{3}+\delta \Bigr)
\end{equation}
\begin{equation} 
\omega=-1,
\qquad
\phi(x^0,x^1)=\frac{\Lambda}{\alpha\beta}e^{\beta x^0}-\frac{\beta}2 x^0+\alpha x^1
\label{sol2full}
,
 \end{equation}
where $\alpha$, $\beta$, $\gamma$, $\delta$ and $\lambda$ are constants.

In the resulting solution, we can set $\alpha=\beta=1$ by scaling transformations, then
\begin{equation}
f_0=\exp{x^0}
,\qquad
\left({t_\mu{}' }\right)^2=(-1)^\mu \Bigl(2 t_\mu{}^3+\lambda t_\mu{}^2+\gamma t_\mu+\delta\Bigr)
,
\end{equation}
\begin{equation}
\omega=-1
,\qquad
\phi(x^0,x^1)=\Lambda e^{x^0}-x^0/2+ x^1
.
\end{equation}
The cosmological constant $\Lambda$ remains arbitrary, the Riemann curvature tensor $R_{ijkl}$ and the Weyl conformal curvature tensor $C_{ijkl}$ do not vanish, and for the scalar curvature $R$ we obtain:
\begin{equation}
R=e^{-x^0},
\qquad
C_{ijkl}\neq 0.
\end{equation}

Thus, a nontrivial solution is obtained for Shapovalov wave spaces of type I in the Brans-Dicke theory with a scalar field depending on null variables, with a cosmological constant and an exponential conformal factor depending on the wave variable. Note that a vacuum solution of this type does not arise in general relativity \cite{Osetrin2020Symmetry} .

\subsection{Scalar field of the form 
$\phi=\phi(x^0,x^2,x^3)$.}

Let's consider the case when the scalar field $\phi$ does not depend on the ''ignored'' null variable $x^1$, on which the metric does not depend either.
The system of BDT equations (\ref{tbd}-\ref{trac}), resolved with respect to higher derivatives, takes the following form:
\begin{equation} \phi_{,00}=\frac{f_0{}'}{f_0}\phi_{,0}+\frac 32\frac{{f_0{}'}^2}{{f_0}^2}-\frac{f_0{}''}{f_0}-(\omega+1)\phi_{,0}{}^2,\label{eq00}\end{equation}
\begin{equation} \phi_{,02}=\phi_{,2}\left(\frac 12\frac{f_0{}'}{f_0}-(\omega+1)\phi_0\right),\end{equation}
\begin{equation} \phi_{,03}=\phi_{,3}\left(\frac 12\frac{f_0{}'}{f_0}-(\omega+1)\phi_0\right),\label{eq03}\end{equation}
\begin{equation} \phi_{,22}=\Lambda Wf_0+\frac{\omega\phi_{,3}{}^2-(\omega+2)\phi_{,2}{}^2}2-\frac{\phi_{,2}t_2{}'+\phi_{,3}t_3{}'}{2W},\label{eq22}\end{equation}
\begin{equation} \phi_{,22}+\phi_{,33}=2\Lambda Wf_0-(\phi_{,2}{}^2+\phi_{,3}{}^2)\end{equation}
\begin{equation} \phi_{,23}=-(\omega+1)\phi_{,2}\phi_{,3}+\frac{\phi_{,2}t_3{}'-\phi_{,3}t_2{}'}{2W},\label{eq23}\end{equation}
\begin{equation} 
(2\Lambda-R)Wf_0+\omega(\phi_{,2}{}^2+\phi_{,3}{}^2)=0,
\end{equation}
\begin{equation}
 R=\frac{{t_2{}'}^2+{t_3{}'}^2+W(t_2{}''-t_3{}'')}{W^3f_0},
\label{eq01}
 \end{equation}
\begin{equation} \Lambda(\omega+1)=0.\end{equation}

The study of the compatibility of the subsystem of equations (\ref{eq00}-\ref{eq03}) leads to relations that are further used to classify possible solutions:
\begin{equation} \phi_{,2}(2f_0f_0{}''-3{f_0{}'}^2)=\phi_{,3}(2f_0f_0{}''-3{f_0{}'}^2)=0\label{sovm}.\end{equation}
The compatibility equations (\ref{sovm}) lead to the following two possible cases:
\begin{enumerate}
\item[1)] ${\phi_{,2}}\,{\phi_{,3}}\neq 0,\quad 2f_0f_0{}''-3{f_0{}'}^2=0$;
\item[2)] $\phi_{,2}=\phi_{,3}=0$, \quad $\phi=\phi(x^0)$.
\end{enumerate}
In the following, we will consider these cases separately.

\subsubsection{Scalar field of the form  $\phi=\phi(x^0,x^2,x^3)$ and ${\phi_{,2}}\,{\phi_{,3}}\neq 0$.}

From the compatibility conditions we get $f_0=1/{\left({x^0}\right)^2}$ and now we can integrate the equations (\ref{eq00}-\ref{eq03}). In this case, two cases are possible depending on the values of the constant $\omega$:
\begin{enumerate}
\item[A)] $\omega=-1$, $\phi={ \xi(x^2,x^3)}/{x^0}$.
Substitution into the equation (\ref{eq01}) allows us to separate the coefficients at $x^0$. As a result, we get $R=0$, which, in view of (\ref{riman}) and (\ref{sovm}), means the degenerate case, since in this case the spacetime becomes flat.
\item[B)] $\omega\neq -1$, $\Lambda=0$, $\phi=\frac1{\omega+1}\ln\frac{ \xi(x^2,x^3)}{x^0}$, 
where the function $\xi(x^2,x^3)$ must satisfy the following equations:
\begin{equation} 
\xi_{,22}=\frac{\omega}{2(\omega+1)}\frac{\xi_{,2}{}^2+\xi_{,3}{}^2}{\xi}-\frac{\xi_{,2}t_2{}'+\xi_{,3}t_3{}'}{2W}
\label{Eq22}
,\end{equation}
\begin{equation}
  \xi_{,33}=\frac{\omega}{2(\omega+1)}\frac{\xi_{,2}{}^2+\xi_{,3}{}^2}{\xi}+\frac{\xi_{,2}t_2{}'+\xi_{,3}t_3{}'}{2W}
\label{Eq33}
,
\end{equation}
\begin{equation} 
\xi_{,23}=\frac{\xi_{,2}t_3{}'-\xi_{,3}t_2{}'}{2W},
\label{eq23}
\end{equation}
\begin{equation}
\frac{\omega}{(\omega+1)^2}\frac{\xi_{,2}{}^2+\xi_{,3}{}^2}{\xi^2}=
\frac{{t_2{}'}^2+{t_3{}'}^2+W(t_2{}''-t_3{}'')}{W^2}
.
\label{EqSovm}
\end{equation}
\end{enumerate}

Let's consider the solution when the scalar field $\phi$ admits separation of variables:
\begin{equation}
\xi(x^2,x^3)=\xi_2(x^2)\,\xi_3(x^3)
,\qquad
{{\xi_2}'}\,{{\xi_3}'}\ne 0
.
\label{xi23}
\end{equation}
From here we get the relations:
\begin{equation}
\xi_{,2}=\xi\,
\frac{d}{dx^2}
\log{\xi_2}
,\qquad
\xi_{,3}=\xi\,
\frac{d}{dx^3}
\log{\xi_3}
.
\label{xi2andxi3}
\end{equation}
Substituting the relations (\ref{xi2andxi3}) into the equation (\ref{eq23}) and separating the variables, we get
\begin{equation}
{\xi_2}t_2{}'/{\xi_2}'-2t_2=
{\xi_3}t_3{}'/{\xi_3}'-2t_3
=\mbox{const}=2c
,
\end{equation}
In this way
\begin{equation}
2\,\frac{d\log{{\xi_2}}}{dx^2}=\frac{d\log{(t_2+c)}}{dx^2}
,\qquad
2\,\frac{d\log{{\xi_3}}}{dx^3}=\frac{d\log{(t_3+c)}}{dx^3}
.
\end{equation}
From this we obtain relations relating the functions $t_\mu$ and $\xi_\mu$ of the following form:
\begin{equation}
{\left(\xi_2\right)}^2=a(t_2+c),
\qquad
{\left(\xi_3\right)}^2=b(t_3+c)
\label{Solutions_xi2_and_xi3}
,
\end{equation}
where $a$, $b$ and $c$ are constants.

Since the functions $t_2, t_3$ enter the metric only as a difference, and the constants $a, b$ add only a constant term to the scalar field $\phi$, then the constants $a,b,c$ can be chosen in the simplest way. As a result, we have an intermediate result for the scalar field $\phi$:
\begin{equation} \xi = \sqrt{t_2 t_3},\ \ \ \ \phi=\frac 1{2(\omega+1)}\ln \frac{t_2t_3}{{x^0}^2}.\label {xi_t2t3}\end{equation}
Having this, excluding the second derivatives from the equations (\ref{Eq22}), (\ref{Eq33}), (\ref{EqSovm}), we obtain the following relation
\begin{equation} \omega(\omega+2)=0.\end{equation}
The case $\omega=0$ leads to a flat solution, it is easy to see this, taking into account the relations (\ref{riman}), (\ref{eq01}), thus,
\begin{equation} \omega =-2.\label{omega-2}\end{equation}

The sum of equations (\ref{Eq22}) and (\ref{Eq33}) after substituting (\ref{xi_t2t3}) and (\ref{omega-2}) allows us to separate variables, and we get
\begin{equation} \frac{t_2{}''}{t_2}-\frac 32\frac{{t_2{}'}^2}{{t_2}^2}=-\left(\frac{t_3{}'' }{t_3}-\frac 32\frac{{t_3{}'}^2}{{t_3}^2}\right)=\mbox{const}=\lambda,\end{equation}
 the solution is the following
\begin{equation} 
t_2=\alpha / \cos^2 \sqrt{\lambda} x^2,\ \ \ \ \ t_3=\beta/\cosh^2\sqrt{\lambda} x^3.
\end{equation}
Substituting this solution into the equations (\ref{Eq22}) and (\ref{Eq33}) leads to a simple condition $\alpha=\beta $, further,
scaling constants $\alpha, \lambda$ can be converted to the simplest form, as a result, the final solution has the following form
\begin{equation}
f_0=1/{\left({x^0}\right)^2}
,\qquad
t_2=\epsilon/\cos^2 x^2
,\qquad
t_3=\epsilon/\cosh^2 x^3
,\qquad
\epsilon=\pm 1,
\end{equation}
\begin{equation}
\phi=\ln(x^0\cos x^2\cosh x^3)
,\qquad
\Lambda=0
\label{BDTsolutionII}
.\end{equation}
For the resulting exact solution (\ref{BDTsolutionII}) of the equations of the Brans-Dicke theory, the Riemann curvature tensor, the scalar curvature, and the Weyl conformal curvature tensor do not vanish:
\begin{equation} R=2\epsilon {x^0}^2,\ \ \ \ \ R_{2323}\neq 0,\ \ \ \ \ C_{ijkl}\neq 0.\end{equation}

\subsubsection{Scalar field of the form $\phi=\phi(x^0)$.}

Let's consider the case where the scalar field $\phi$ depends only on the non-ignorable wave variable $x^0$
($\phi_{,2}=\phi_{,3}=0$), on which the wave metric itself depends.
Equations (\ref{eq22}) and (\ref{eq01}) imply that
the cosmological constant $\Lambda$ and the scalar curvature $R$ vanish in this case:
\begin{equation} \Lambda=R=0.\end{equation}
Then you can immediately separate the variables in the equation (\ref{eq01}), which allows you to find
differential equations for the functions $t_\mu(x^\mu)$:
\begin{equation} {\left(t_{2}{}'\right)}^2=a {t_{2}}^2+2b t_{2}+c,
\label{t23Casex0}
\end{equation}
\begin{equation} {\left(t_{3}{}'\right)}^2=-\Bigl(a {t_{3}}^2+2b t_{3}+c\Bigr),
\end{equation}
where $a$, $b$ and $c$ are constants.

The system of field equations is left with the only equation (\ref{eq00}) connecting the function $f_0$ (scale factor of the metric) and the scalar field $\phi(x^0)$:
\begin{equation} \phi''=\phi'\left(\frac {f_0{}'}{f_0}-(\omega+1)\phi'\right)-\frac{2f_0f_0{}''-3{f_0 {}'}^2}{2{f_0}^2}\label{eqfi}
.\end{equation}
The equation (\ref{eqfi}) obviously admits solutions for the scalar field $\phi=\phi_0(x^0)$ for a given conformal factor $f_0(x^0)$ of the metric.

In the case under consideration, the Weil tensor $C_{ijkl}$ vanishes and only two nonzero components of the Riemann curvature tensor remain
$R_{ijkl}$:
\begin{equation} R_{0202}=R_{0303}=\frac{W}{4f_0}(3{f_0{}'}^2-2f_0f_0{}'')
,\end{equation}
\begin{equation}
C_{ijkl}
=R_{2323}
=0.
\end{equation}
Thus, this solution is the case of a conformally flat space, and the metric and components of the Riemann curvature tensor depend on the wave variable $x^0$.
The scalar field $\phi$ and the conformal factor of the metric $f_0$ remain arbitrary functions of the wave variable $x^0$, but are related by the equation (\ref{eqfi}).

Note that the equation (\ref{eqfi}) admits a de Sitter-type solution.
Consider the scale factor of the metric in the following form:
\begin{equation}
f_0(x^0)=k e^{\beta x^0}
,\qquad
k,\beta - \mbox{const}
\label{f0DeSitter}
.\end{equation}

Then the equation (\ref{eqfi}) takes the following form:
\begin{equation} \phi''=\phi'\left(\beta-(\omega+1)\phi'\right)+\beta^2/2
\label{eqfifi}
.\end{equation}
Equation (\ref{eqfifi}) for scalar field $\phi$ for metric (\ref{metric1}) with scale factor (\ref{f0DeSitter})
has three solutions depending on the values of the constant
$\omega$:
\begin{enumerate}
\item[1)] 
$\omega=-1$. Then
\begin{equation} \phi=\beta e^{\beta x^0}-\frac{\beta}2 x^0,\end{equation}
\item[2)]
$\omega<-3/2$. Then
\begin{equation} \phi=\frac{\beta}{2(\omega+1)}x^0+\frac 1{\omega+1}\ln\cos(\beta\sqrt{-(2\omega+3)}\,x^0/2),\end{equation}
\item[3)]
$\omega>-3/2, \ \ \omega\neq-1$. Then
\begin{equation} \phi=\frac{\beta}{2(\omega+1)}x^0-\frac 1{\omega+1}\ln\cosh(\beta\sqrt{2\omega+3}\,x^0/2)
,
\end{equation}
\end{enumerate}
where $\beta$ is a constant parameter.

Thus, we have obtained a number of exact solutions to the equations of the scalar-tensor Brans-Dicke theory in vacuum for the type I Shapovalov wave space. Recall that there are no exact solutions of this type for the case of vacuum Einstein equations \cite{Osetrin2020Symmetry}, since Shapovalov spaces of type I degenerate in the case of vacuum Einstein equations.

Thus, scalar-tensor theories of gravity can give such types of exact models of gravitational waves that do not arise in general relativity.

\section{
Equations of motion and trajectories of test particles
}

Let's consider the equation of motion of test particles in a gravitational field in the Hamilton-Jacobi formalism (the speed of light is chosen to be unity):
$$
g^{\alpha\beta}\frac{\partial S}{\partial x^\alpha}\frac{\partial S}{\partial x^\beta}=m^2
.
$$

The complete integral of the Hamilton-Jacobi equation (\ref{HJE}) for the Shapovalov spacetimes in the privileged coordinate system can be represented in a separated form:
\begin{equation}
S={\theta}_0(x^0)+{\theta}_1(x^1)+{\theta}_2(x^2)+{\theta}_3(x^3),
\end{equation}
moreover, for the ignored variable $x^1$, on which the metric in the privileged coordinate system does not depend, the function ${\theta}_1(x^1)$ can be reduced by admissible coordinate transformations to the form $kx^1$, where $k$ is a constant.

From the equation (\ref{HJE}), separating the variable $x^0$, we get:
\begin{equation}
\frac{m^2}{f_0(x^0)}-2k{{\theta}_0}'(x^0)=\frac{1}{t_3(x^3)-t_2(x^2)}
\Bigl(
\left( {{\theta}_2}'(x^2) \right)^2
+
\left( {{\theta}_3}'(x^3) \right)^2
\Bigr)
=\mbox{const}=p
.
\end{equation}
Finally, separating the variables $x^2$ and $x^3$ we get:
\begin{equation}
\left( {{\theta}_2}'(x^2) \right)^2
+p t_2(x^2)
=
-\left( {{\theta}_3}'(x^3) \right)^2
+p t_3(x^3)
=\mbox{const}=q
.
\end{equation}

Thus, for the complete integral of the particle action function $S(x^\alpha,k,p,q)$ we obtain the following expression:
$$
S=kx^1-\frac{p}{2k}x^0+\frac{m^2}{2k}\int{\frac{dx^0}{f_0(x^0)}}
+\varepsilon_2\int{\sqrt{q-pt_2(x^2)}\,dx^2}
$$
\begin{equation}
\mbox{}
+\varepsilon_3\int{\sqrt{pt_3(x^3)-q}\,dx^2}
,\qquad
\varepsilon_2,\varepsilon_3=\pm 1
,
\end{equation}
where $k$, $p$ and $q$ are constant parameters.

The equations of trajectories of test particles for the considered Shapovalov spaces of type I in the Hamilton-Jacobi formalism take the form:
\begin{equation}
\frac{\partial S}{\partial k}
=\mbox{const}=\sigma_1
\quad\to\quad
x^1+\frac{p}{2k^2}x^0+\frac{m^2}{2k^2}\int{\frac{dx^0}{f_0(x^0)}}=\sigma_1
,\qquad
k\ne 0
\label{Trajectory1}
,
\end{equation}
\begin{equation}
\frac{\partial S}{\partial p}
=\mbox{const}=\sigma_2
\quad\to\quad
-\frac{x^0}{2k}
-\frac{\varepsilon_2}{2}\int{\frac{t_2(x^2)\,dx^2}{\sqrt{q-pt_2(x^2)}}}
+\frac{\varepsilon_3}{2}\int{\frac{t_3(x^3)\,dx^3}{\sqrt{pt_3(x^3)-q}}}
=\sigma_2
,
\label{Trajectory2}
\end{equation}
\begin{equation}
\frac{\partial S}{\partial q}
=\mbox{const}=\sigma_3
\quad\to\quad
\frac{\varepsilon_2}{2}\int{\frac{dx^2}{\sqrt{q-pt_2(x^2)}} }
-
\frac{\varepsilon_3}{2}\int{\frac{dx^3}{\sqrt{pt_3(x^3)-q}}}
=\sigma_3
,
\label{Trajectory3}
\end{equation}
where $\sigma_1$, $\sigma_2$ and $\sigma_3$ are new independent constant parameters of the test particle motion determined by the initial conditions. Thus, the motion of test particles in the considered space-time models is determined by a set of constant parameters $k$, $p$, $q$, $\sigma_1$, $\sigma_2$ and $\sigma_3$, given by the initial conditions.

Note that the proper time of the test particle $\tau$ using the obtained trajectory equations can be written in the following form in terms of the coordinate variables on the trajectory:
\begin{equation}
\tau=S\,\bigr |_{m=1}=2k x^1+\frac{p}{k}\,x^0
.
\label{taux0x1}
\end{equation}
Here we set the additive constant equal to zero by choosing the origin. Thus proper time, as expected, is a linear combination of the null variables $x^0$ and $x^1$.

If we substitute (\ref{taux0x1}) into the equation (\ref{Trajectory1}), then on the trajectories of test particles we get:
\begin{equation}
\tau=
-
\frac{1}{k}\int{\frac{dx^0}{f_0(x^0)}}+2k\sigma_1
\label{taux0}
,
\end{equation}

Thus, the equations (\ref{taux0x1}), (\ref{taux0}) together with the equations (\ref{Trajectory2}) and (\ref{Trajectory3}) define the coordinates of the test particle $x^\alpha$ on the trajectory as a function of of the proper time of the particle $\tau$.

\section{Conclusion}
The exact solutions of the vacuum equations of the Brans-Dicke theory of gravity for Shapovalov type I wave spaces are found. Gravitational-wave solutions are obtained, depending on the maximum possible number of variables for wave metrics of 4-dimensional space-time
(three variables, including the wave variable) in privileged coordinate systems, where it is possible to separate the wave variablesin the Hamilton-Jacobi equation for test particles and in the eikonal equation for radiation. The situation is shown to differ from the general theory of relativity, where these gravitational-wave models based on Einstein's vacuum equations degenerate \cite{Osetrin2020Symmetry}. Thus, Shapovalov's wave spaces provide an additional mathematical tool for obtaining exact models of gravitational waves, which makes it possible to study differences in modified gravity theories and form tests for observational verification of these differences.

%

\authorcontributions{Conceptualization, K.O.; methodology, K.O. and I.K.; validation, A.F., I.K., E.O. and K.O.;  investigation, K.O., I.K., A.F. and E.O; writing---original draft preparation, K.O.; supervision, K.O.; project administration, K.O.; funding acquisition, K.O. All authors have read and agreed to the published version of the manuscript.}

\funding{The reported study was funded by RFBR, project number N~20-01-00389~A.}

\conflictsofinterest{The authors declare no conflict of interest. }

\institutionalreview{Not applicable.}

\informedconsent{Not applicable.}

\dataavailability{Not applicable.} 

%

\reftitle{References}

\end{document}